\documentclass[aps,twocolumn]{revtex4}

\usepackage{amsfonts,amsmath,amssymb}
\usepackage[dvips]{graphicx}
\usepackage{bm}
\usepackage{multirow}

\DeclareMathOperator{\Tr}{Tr}
\newcommand{\osp}{\ensuremath{\mathfrak{osp} \left( 32 \middle| 1 \right)}}
\begin{document}

\title{Eleven-Dimensional Gauge Theory for the M~Algebra\\%
as an Abelian Semigroup Expansion of \osp}

\date{December 17, 2006}

\author{Fernando Izaurieta}
\email{fizaurie@udec.cl}

\author{Eduardo Rodr\'{\i}guez}
\email{edurodriguez@udec.cl}

\affiliation{Departamento de Matem\'{a}ticas y F\'{\i}sica Aplicadas,
Universidad Cat\'{o}lica de la Sant\'{\i}sima Concepci\'{o}n, Concepci\'{o}n, Chile}

\author{Patricio Salgado}
\email{pasalgad@udec.cl}

\affiliation{Departamento de F\'{\i}sica, Universidad de Concepci\'{o}n, Casilla 160-C, Concepci\'{o}n, Chile}

\begin{abstract}
A new Lagrangian realizing the symmetry of the M~Algebra in eleven-dimensional space-time is presented. By means of the novel technique of Abelian Semigroup Expansion, a link between the M~Algebra and the orthosymplectic algebra \osp\ is established, and an M~Algebra-invariant symmetric tensor of rank six is computed. This symmetric invariant tensor is a key ingredient in the construction of the new Lagrangian. The gauge-invariant Lagrangian is displayed in an explicitly Lorentz-invariant way by means of a subspace separation method based on the extended Cartan homotopy formula.
\end{abstract}


\maketitle

\section{\label{intro}Introduction}

String Theory and eleven-dimensional Supergravity became inextricably linked after the arrival of the M-Theory Paradigm. All efforts notwithstanding, the low-energy regime of M~Theory remains better known than its non-perturbative description. However, the possibility has been pointed out that M~Theory may be non-perturbatively related to, or even formulated as, an eleven-dimensional Chern--Simons theory~\cite{Tro97,Horava,Nastase} (see also~\cite{Bag02}).

Chern--Simons (CS) Theory has quite compelling features. On one hand, it belongs to the restricted class of gauge field theories, with a one-form gauge connection as the sole dynamical field. On the other hand, and in contrast with usual Yang--Mills theory, there's no \textit{a priori} metric needed to define the CS Lagrangian, so that the theory turns out to be background-free. CS Supergravities (see, e.g.,~\cite{Edel06b} and references therein) exist in every odd dimension; three-dimensional General Relativity was famously quantized by making the connection to CS~\cite{Witten}.

In recent times, an even more appealing generalization of this idea has been presented, the so-called Transgression form Lagrangians. Transgression forms~\cite{Polaco1,Polaco2,Nosotros,Mora1,CECS,Nosotros2} are the matrix where CS forms stem from. The main difference between CS and Transgression forms concerns a new, regularizing boundary term which renders the Transgression form fully gauge invariant. As a consequence, the boundary conditions and Noether charges computed from a transgression action have the
chance to be physically meaningful.

Since a gauge field theory for the M~Algebra may take us one step closer to understanding the non-perturbative description of M~Theory, the importance of the formulation of a CS/Transgression form theory for the M-Algebra is clear. A priori, the construction of a CS Supergravity for the M~Algebra would seem something straightforward to do, especially since CS Supergravities for \osp\ are already well-known \cite{Horava,Nastase,Edel06b}. This is however not the case, and the construction is actually highly nontrivial. The reason is that in both cases, for CS and Transgression forms, the key ingredient in the construction is the invariant tensor. And precisely in the case of the non-semisimple M~Algebra, the direct option of using the supertrace as invariant tensor is not a fruitful one.

This problem has been dealt with in Refs.~\cite{Has03,Has05} using a physicist's approach: the Noether method. Starting from the Poincar\'{e} CS Lagrangian, a CS Form for the M~Algebra is recursively constructed, adding new terms to finally reach an invariant Lagrangian. After the Lagrangian is constructed, it is possible to read back the invariant tensor. This approach has proved succesful, but it has some drawbacks: (i) it requires a lot of physicist's insight and cleverness and (ii) as the authors of~\cite{Has03,Has05} make clear, the method does not rule out the possiblity of extra terms in the Lagrangian.

On the other hand, a more matemathical point of view has been developed in Ref.~\cite{Azcarraga Et Al}, where the M~Algebra has been shown to correspond to an \emph{expansion}%
~\footnote{The M~Algebra (with 583 bosonic generators) is sometimes regarded in the literature as a \emph{contraction} of \osp. That this cannot be correct can be seen by observing that a contraction of \osp\ (with only 528 bosonic generators) would lack the 55 generators of the Lorentz automorphism piece, since a contraction cannot change the number of generators. For a thorough discussion of this problem, see Ref.~\cite{Azcarraga Et Al}.}
of \osp. Expansions stand out among other algebra manipulation methods (such as contractions, deformations and extensions) as the only one which is able of changing the dimension of the algebra; in general, it leads to algebras with a dimensionality higher than the original one.

In a nutshell, the expansion method considers the original algebra as described by its associated Maurer--Cartan (MC) forms on the group manifold. Some of the group parameters are rescaled by a factor $\lambda$, and the MC forms are expanded as a power series in $\lambda$. This series is finally truncated in a way that assures the closure of the expanded algebra. The subject is thoroughly treated in Refs.~\cite{Azcarraga Et Al,AzcarragaSuperspace,deAz07,LibroAzcarraga}.

In the expansions approach, the algebra is formulated in terms of the MC forms, and therefore, the CS form for the M~Algebra must be written through a free differential algebra series from the full \osp-CS form. Again, to extract from there an invariant tensor for the M~Algebra proves to be nontrivial.

Both approaches focus on constructing directly the CS form. In this article, a third alternative is considered: the Lie Algebras $S$-expansion method, which focuses on the construction of the \emph{invariant tensor}. This procedure, developed in general in Ref.~\cite{NosExp}, is formulated in terms of the original Lie algebra \emph{generators} and an abelian semigroup $S$. Given this original Lie algebra and the abelian semigroup as inputs, the $S$-expansion method gives as output a new Lie algebra, and besides it, general expressions for the invariant tensor for it in terms of the semigroup structure.

The paper is organized as follows. In sec.~\ref{sexpa} the derivation of the M~Algebra as an abelian semigroup expansion of \osp\ is performed, and a way to construct an M~algebra-invariant tensor is found. Some aspects of the transgression Lagrangian are reviewed in Section~\ref{malglag}, where use of the subspace separation method produces a new explicit action for an eleven-dimensional transgression gauge field theory. In sec.~\ref{dyna} we comment on the dynamics produced by the transgression Lagrangian. We
close with conclusions and some final remarks in sec.~\ref{final}.

\section{\label{sexpa}The M~Algebra as an $S$-Expansion of \osp}

In this section we briefly review the general method of abelian semigroup expansion and its application in obtaining the M~Algebra as an $S$-Expansion of \osp. We refer the reader to~\cite{NosExp} for the details.

\subsection{\label{sexpapro}The $S$-Expansion Procedure}

Consider a Lie algebra $\mathfrak{g}$ and a finite abelian semigroup $S = \left\{ \lambda_{\alpha} \right\}$. According to Theorem~3.1 from Ref.~\cite{NosExp}, the direct product $S \times \mathfrak{g}$ is also a Lie algebra. Interestingly, there are cases when it is possible to systematically extract subalgebras from $S \times \mathfrak{g}$. Start by decomposing $\mathfrak{g}$ in a direct sum of subspaces, as in
\begin{equation}
\mathfrak{g} = \bigoplus_{p \in I} V_{p},
\end{equation}
where $I$ is a set of indices. The internal subspace structure of $\mathfrak{g}$ can be codified through~\footnote{Here $2^{I}$ denotes the set of all subsets of $I$.} the mapping $i:I\times I\rightarrow 2^{I}$, where the subsets $i\left( p,q \right) \subset I$ are such that
\begin{equation}
\left[ V_{p}, V_{q} \right] \subset \bigoplus_{r \in i\left( p,q \right)} V_{r}.
\end{equation}
When the semigroup $S$ can be decomposed in subsets $S_{p}$, $S = \bigcup_{p \in I} S_{p}$, such that they satisfy the condition%
~\footnote{Here $S_{p} \cdot S_{q} \subset S$ is defined as the set which includes all products between all elements from $S_{p}$ and all elements from $S_{q}$.}
\begin{equation}
S_{p} \cdot S_{q} \subset \bigcap_{r \in i\left( p,q \right)} S_{r}, \label{ResonantCondition}
\end{equation}
then we have that
\begin{equation}
\mathfrak{G}_{\text{R}} = \bigoplus_{p \in I} S_{p} \times V_{p}
\end{equation}
is a `resonant subalgebra' of $S \times \mathfrak{g}$ (see Theorem~4.2 from Ref.~\cite{NosExp}).

An even smaller algebra can be obtained when there is a zero element in the semigroup, i.e., an element $0_{S} \in S$ such that, for all $\lambda_{\alpha} \in S$, $0_{S} \lambda_{\alpha} = 0_{S}$. When this is the case, the whole $0_{S} \times \mathfrak{g}$ sector can be removed from the resonant subalgebra by imposing $0_{S} \times \mathfrak{g} = 0$. The remaining piece, to which we refer to as $0_{S}$-reduced algebra, continues to be a Lie algebra (for a proof of this fact and some more general cases, see $0_{S}$-reduction and Theorem~6.1 from Ref.~\cite{NosExp}).

In the next section these mathematical tools will be used in order to show how the M~algebra can be constructed from \osp.

\subsection{\label{malgsexpa}M~Algebra as an $S$-expansion}

In this section we roughly sketch the steps to be undertaken in order to
obtain the M~algebra as an $S$-Expansion of \osp.

As with any expansion, the first step consists in splitting the \osp\ algebra in distinct subspaces. This is accomplished by defining
\begin{align}
V_{0}  &  =\left\{  \bm{J}_{ab}^{\left(  \mathfrak{osp}\right)
}\right\}  ,\\
V_{1}  &  =\left\{  \bm{Q}^{\left(  \mathfrak{osp}\right)  }\right\}
,\\
V_{2}  &  =\left\{  \bm{P}_{a}^{\left(  \mathfrak{osp}\right)
},\bm{Z}_{a_{1}\cdots a_{5}}^{\left(  \mathfrak{osp}\right)
}\right\}  .
\end{align}
Here $V_{0}$ corresponds to the Lorentz algebra, $V_{1}$ to the fermions and
$V_{2}$ to the remaining bosonic generators, namely AdS boosts and the
M5-brane piece. The algebraic structure satisfied by these subspaces is common
to every superalgebra, as can be seen from the equations
\begin{align}
\left[  V_{0},V_{0}\right]   &  \subset V_{0},\label{v0v0v0}\\
\left[  V_{0},V_{1}\right]   &  \subset V_{1},\\
\left[  V_{0},V_{2}\right]   &  \subset V_{2},\\
\left[  V_{1},V_{1}\right]   &  \subset V_{0}\oplus V_{2},\\
\left[  V_{1},V_{2}\right]   &  \subset V_{1},\\
\left[  V_{2},V_{2}\right]   &  \subset V_{0}\oplus V_{2}. \label{v2v2v0v2}%
\end{align}

The second step is particular to the method of $S$-expansions, and deals with
finding an abelian semigroup $S$ which can be partitioned in a `resonant' way
with respect to (\ref{v0v0v0})--(\ref{v2v2v0v2}). This semigroup exists and is
given by $S_{\text{E}}^{\left(  2\right)  }=\left\{  \lambda_{0},\lambda
_{1},\lambda_{2},\lambda_{3}\right\}  $, with the defining product%
\begin{equation}
\lambda_{\alpha}\lambda_{\beta}=\left\{
\begin{array}[c]{ll}
  \lambda_{\alpha+\beta}, & \text{when } \alpha+\beta\leq2, \\
  \lambda_{3}, & \text{otherwise.}
\end{array}
\right.
\label{SE(N)Product}
\end{equation}

A straightforward but important observation is that, for each
$\lambda_{\alpha}\in S_{\text{E}}^{\left(  2\right)  }$,
$\lambda_{3}\lambda_{\alpha}=\lambda_{3}$,
so that $\lambda_{3}$ plays the r\^{o}le of the zero
element inside $S_{\text{E}}^{\left(  2\right)  }$.

Consider now the partition
$S_{\text{E}}^{\left(  2\right)  }=S_{0}\cup S_{1}\cup S_{2}$, with
\begin{align}
S_{0}  &  =\left\{  \lambda_{0},\lambda_{2},\lambda_{3}\right\}
,\label{ResS0}\\
S_{1}  &  =\left\{  \lambda_{1},\lambda_{3}\right\}  ,\label{ResS1}\\
S_{2}  &  =\left\{  \lambda_{2},\lambda_{3}\right\}  . \label{ResS2}%
\end{align}
This partition is said to be resonant, since it satisfies [compare eqs.~(\ref{v0v0v0})--(\ref{v2v2v0v2}) with eqs.~(\ref{SuperS0S0=S0})--(\ref{SuperS2S2=S0nS2})]
\begin{align}
S_{0}\cdot S_{0}  &  \subset S_{0},\label{SuperS0S0=S0}\\
S_{0}\cdot S_{1}  &  \subset S_{1},\label{SuperS0S1=S1}\\
S_{0}\cdot S_{2}  &  \subset S_{2},\label{SuperS0S2=S2}\\
S_{1}\cdot S_{1}  &  \subset S_{0}\cap S_{2},\label{SuperS1S1=S0nS2}\\
S_{1}\cdot S_{2}  &  \subset S_{1},\label{SuperS1S2=S1}\\
S_{2}\cdot S_{2}  &  \subset S_{0}\cap S_{2}. \label{SuperS2S2=S0nS2}%
\end{align}

Theorem~4.2 from Ref.~\cite{NosExp} now assures us that
\begin{equation}
\mathfrak{G}_{\text{R}}=\left(  S_{0}\times V_{0}\right)  \oplus\left(
S_{1}\times V_{1}\right)  \oplus\left(  S_{2}\times V_{2}\right)
\label{ResonantePreM}%
\end{equation}
is a \emph{resonant subalgebra} of $S_{\text{E}}^{\left( 2 \right) } \times \mathfrak{g}$.

\begin{table}
\caption{\label{tab:sxv}The M~algebra can be regarded as an $S_{\text{E}}^{(2)}$-Expansion of \osp. The table shows the relation between generators from both algebras. The three levels correspond to the three columns in Fig.~\ref{FigMAlgebra} or, alternatively, to the three subsets into which $S_{\text{E}}^{(2)}$ has been partitioned.}
\begin{ruledtabular}
\begin{tabular}[c]{cccrclc}
\hspace{.10\columnwidth} &
$\mathfrak{G}_{\text{R}}$ Subspaces &
\hspace{.17\columnwidth} &
\multicolumn{3}{c}{Generators} &
\hspace{.10\columnwidth} \\
\hline &
\multirow{3}*{$S_{0} \times V_{0}$}
  & &
  $\bm{J}_{ab}$
  & $=$ &
  $\lambda_{0} \bm{J}_{ab}^{\left(  \mathfrak{osp} \right) }$
  & \\ &
  & &
  $\bm{Z}_{ab}$
  & $=$ &
  $\lambda_{2} \bm{J}_{ab}^{\left(  \mathfrak{osp} \right) }$
  & \\ &
  & &
  $\bm{0}$
  & $=$ &
  $\lambda_{3} \bm{J}_{ab}^{\left( \mathfrak{osp} \right) }$
  & \\
\hline &
\multirow{2}*{$S_{1} \times V_{1}$}
  & &
  $\bm{Q}$
  & $=$ &
  $\lambda_{1} \bm{Q}^{\left(  \mathfrak{osp} \right) }$
  & \\ &
  & &
  $\bm{0}$
  & $=$ &
  $\lambda_{3} \bm{Q}^{\left( \mathfrak{osp} \right) }$ \\
\hline &
\multirow{4}*{$S_{2} \times V_{2}$}
  & &
  $\bm{P}_{a}$
  & $=$ &
  $\lambda_{2} \bm{P}_{a}^{\left(  \mathfrak{osp} \right) }$
  & \\ &
  & &
  $\bm{Z}_{abcde}$
  & $=$ &
  $\lambda_{2} \bm{Z}_{abcde}^{\left(  \mathfrak{osp} \right) }$
  & \\ &
  & &
  $\bm{0}$
  & $=$ &
  $\lambda_{3} \bm{P}_{a}^{\left( \mathfrak{osp} \right) }$
  & \\ &
  & &
  $\bm{0}$
  & $=$ &
  $\lambda_{3} \bm{Z}_{abcde}^{\left( \mathfrak{osp} \right) }$
\end{tabular}
\end{ruledtabular}
\end{table}

As a last step, impose the condition $\lambda_{3}\times\mathfrak{g}=0$ on
$\mathfrak{G}_{\text{R}}$ and relabel its generators as in Table~\ref{tab:sxv}. This procedure gives us the M~algebra, whose (anti)commutation relations are recalled in Table~\ref{tab:MAlg}.

\begin{table}
\caption{\label{tab:MAlg}(Anti)commutation relations for the M algebra.
Here $\Gamma_{a}$ are Dirac matrices in $d=11$.}
\begin{ruledtabular}
\begin{align}
\left[  \bm{J}^{ab},\bm{J}_{cd}\right] & = \delta
_{ecd}^{abf}\bm{J}_{\phantom{e}f}^{e}, \\
\left[  \bm{J}^{ab},\bm{P}_{c}\right] & = \delta_{ec}%
^{ab}\bm{P}^{e},\\
\left[  \bm{J}^{ab},\bm{Z}_{cd}\right] & = \delta
_{ecd}^{abf}\bm{Z}_{\phantom{e}f}^{e},\\
\left[  \bm{J}^{ab},\bm{Z}_{c_{1}\cdots c_{5}}\right]
& = \frac{1}{4!}\delta_{dc_{1}\cdots c_{5}}^{abe_{1}\cdots e_{4}}\bm{Z}%
_{\phantom{d}e_{1}\cdots e_{4}}^{d}, \\
\left[  \bm{J}_{ab},\bm{Q}\right] & = -\frac{1}{2}%
\Gamma_{ab}\bm{Q}, \\
\left[  \bm{P}_{a},\bm{P}_{b}\right] & = \bm{0},\\
\left[  \bm{P}_{a},\bm{Z}_{bc}\right] & = \bm{0}, \\
\left[  \bm{P}_{a},\bm{Z}_{b_{1}\cdots b_{5}}\right]
& = \bm{0},\\
\left[  \bm{Z}_{ab},\bm{Z}_{cd}\right] & = \bm{0},\\
\left[  \bm{Z}_{ab},\bm{Z}_{c_{1}\cdots c_{5}}\right]
& = \bm{0},\\
\left[  \bm{Z}_{a_{1}\cdots a_{5}},\bm{Z}_{b_{1}\cdots b_{5}%
}\right] & = \bm{0}, \\
\left[  \bm{P}_{a},\bm{Q}\right] & = \bm{0}, \label{PQM} \\
\left[  \bm{Z}_{ab},\bm{Q}\right] & = \bm{0},\\
\left[  \bm{Z}_{abcde},\bm{Q}\right] & = \bm{0}, \label{PZ5M}
\end{align}
\begin{equation}
\left\{  \bm{Q},\bar{\bm{Q}}\right\} = \frac{1}{8}\left(
\Gamma^{a}\bm{P}_{a}-\frac{1}{2}\Gamma^{ab}\bm{Z}_{ab}%
+\frac{1}{5!}\Gamma^{abcde}\bm{Z}_{abcde}\right)  .
\end{equation}
\end{ruledtabular}
\end{table}

\begin{figure}
 \includegraphics[width=.55\columnwidth]{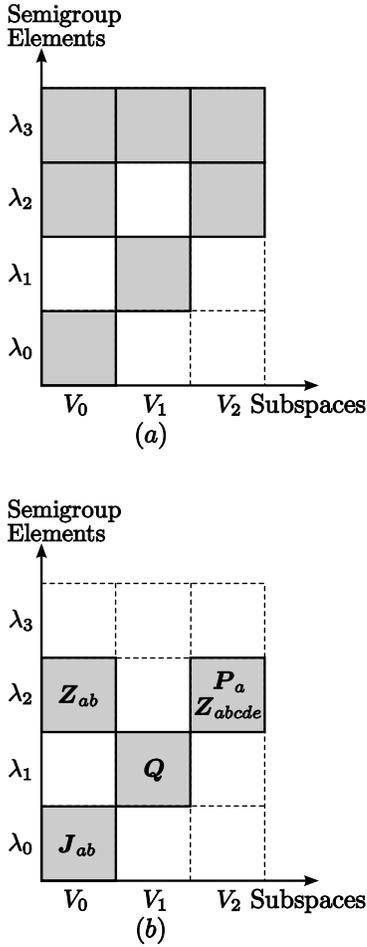}
 \caption{\label{FigMAlgebra} (\textit{a}) The shaded region denotes the resonant subalgebra $\mathfrak{G}_{\text{R}}$. (\textit{b}) Shaded areas correspond to the M~algebra itself, which is obtained from $\mathfrak{G}_{\text{R}}$ by mapping the $\lambda_{3} \times \osp$ sector to zero.}
\end{figure}

A clearer picture of the algebra's structure can be obtained from the diagram in Fig.~\ref{FigMAlgebra}. The subspaces of \osp\ are represented on the horizontal axis, and the semigroup elements on the vertical one. The shaded region on the left corresponds to the resonant subalgebra, including the
$\lambda_{3} \times \osp$
sector, which is mapped to zero via the $0_{S}$-reduction. The gray sector on the right corresponds to the M~algebra itself. The diagram allows us to graphically encode the subset partition (\ref{ResS0})--(\ref{ResS2}) on each column, and makes checking the closure of the algebra a straightforward matter.

Large sectors of the resonant subalgebra are abelianized after imposing the condition $\lambda_{3}\times \osp = 0$. This condition also plays a fundamental r\^{o}le in the shaping of the invariant tensor for the M~algebra as an $S$-expansion of \osp. In this way, its effects are felt all the way down to the theory's specific dynamic properties.

\subsection{\label{malginvten}M-Algebra Invariant Tensor}

Finding all possible invariant tensors for an \emph{arbitrary} algebra remains, to the best of our knowledge, as an important open problem. Nevertheless, once a matrix representation for a Lie algebra is known, the (super)trace always provides with an invariant tensor. But precisely in our case, this is not a wise choice: in general, it is possible to prove that when the condition $0_{S}\times\mathfrak{g}=0$ is imposed, the supertrace for the $S$-expanded algebra generators will correspond to just a very small piece of the whole (super)trace for the $\mathfrak{g}$-generators. For the particular case of the M~algebra, the only non-vanishing component of the supertrace is
$\Tr \left( \bm{J}_{a_{1} b_{1}} \cdots \bm{J}_{a_{n} b_{n}} \right)$.
A CS Lagrangian constructed with this invariant tensor would lead to an `exotic gravity', where the fermions, the central charges and even the vielbein would be absent from the invariant tensor. For this reason, it becomes a necessity to work out other kinds of invariant tensors; very interesting work on precisely this point has been developed in Refs.~\cite{Has03,Has05}, where an invariant tensor for the M~algebra is obtained from the Noether method, finally leading to a CS M-algebra Supergravity in eleven dimensions.

In the context of an $S$-expansion, Theorems~7.1 and~7.2 from Ref.~\cite{NosExp} provide with non-trivial invariant tensors different from the supertrace.

Let $\lambda _{\alpha _{1}},\ldots ,\lambda _{\alpha _{n}}\in S$ be
arbitrary elements of the semigroup $S$. Their product can be written as
\begin{equation}
\lambda _{\alpha _{1}}\cdots \lambda _{\alpha_{n}}=\lambda_{\gamma \left(
\alpha _{1},\ldots ,\alpha _{n}\right) }.
\end{equation}%
This product law can be conveniently encoded by the \emph{$n$-selector}
$K_{\alpha_{1} \cdots \alpha_{n}}^{\phantom{\alpha_{1} \cdots \alpha_{n}} \rho}$,
which is defined as
\begin{equation}
K_{\alpha _{1}\cdots \alpha_{n}}^{\phantom{\alpha_{1} \cdots \alpha_{n}} \rho }=\left\{
\begin{array}{l}
  1 \text{, when } \rho =\gamma \left( \alpha_{1},\ldots ,\alpha_{n}\right), \\
  0 \text{, otherwise.}
\end{array}
\right.
\end{equation}

Theorem~7.1 from Ref.~\cite{NosExp} states that
\begin{equation}
\left\langle \bm{T}_{\left( A_{1},\alpha _{1}\right) }\cdots \bm{T}_{\left(
A_{n},\alpha _{n}\right) }\right\rangle =\alpha _{\gamma }K_{\alpha
_{1}\cdots \alpha _{n}}^{\phantom{\alpha_{1} \cdots \alpha_{n}} \gamma }\left\langle \bm{T}_{A_{1}}\cdots %
\bm{T}_{A_{n}}\right\rangle   \label{teninvg}
\end{equation}
corresponds to an invariant tensor for the $S$-expanded algebra without $0_{S}$-reduction,
where $\alpha _{\gamma }$ are arbitrary constants.

When the semigroup contains a zero element $0_{S}\in S$, a smaller algebra
can be obtained by `$0_{S}$-reducing' the $S$-expanded algebra, i.e., by
mapping all elements of the form $0_{S}\times \mathfrak{g}$ to zero.
Writing $\lambda _{i}$ for the nonzero elements of $S$, Theorem~7.2 from Ref.~\cite{NosExp} assures that
\begin{equation}
\left\langle \bm{T}_{\left( A_{1},i_{1}\right) }\cdots \bm{T}_{\left(
A_{n},i_{n}\right) }\right\rangle =\alpha _{j}K_{i_{1}\cdots i_{n}}^{\phantom{i_{1} \cdots i_{n}} j}\left\langle \bm{T}_{A_{1}}\cdots \bm{T}_{A_{n}}\right\rangle
\label{<T...T>=aK<T...T>}
\end{equation}%
is an invariant tensor for the $0_{S}$-reduced algebra, with $\alpha _{j}$
being arbitrary constants. As can be seen by comparing eq.~(\ref{teninvg})
with eq.~(\ref{<T...T>=aK<T...T>}), this invariant tensor corresponds to a
`pruning' of~(\ref{teninvg}).

In the M-algebra case, one must compute the components of
$K_{i_{1} \cdots i_{6}}^{\phantom{i_{1} \cdots i_{6}} j}$
for $S_{\text{E}}^{\left( 2\right) }$. Using the
multiplication law~(\ref{SE(N)Product}), these are easily seen to be
\begin{equation}
K_{i_{1} \cdots i_{6}}^{\phantom{i_{1} \cdots i_{6}} j} =
\delta_{i_{1} + \cdots + i_{6}}^{j},
\label{n-selector=delta}
\end{equation}
where $\delta$ is the Kronecker delta.

Using eqs.~(\ref{<T...T>=aK<T...T>}) and~(\ref{n-selector=delta}), we have that the \emph{only} non-vanishing components of the M~algebra-invariant tensor are given by
\begin{align}
\left\langle \bm{J}_{a_{1}b_{1}}\cdots \bm{J}_{a_{6}b_{6}}\right\rangle _{%
\text{M}}& =\alpha _{0}\left\langle \bm{J}_{a_{1}b_{1}}\cdots \bm{J}%
_{a_{6}b_{6}}\right\rangle _{\mathfrak{osp}},  \label{itmalg1} \\
\left\langle \bm{J}_{a_{1}b_{1}}\cdots \bm{J}_{a_{5}b_{5}}\bm{P}%
_{c}\right\rangle _{\text{M}}& =\alpha _{2}\left\langle \bm{J}%
_{a_{1}b_{1}}\cdots \bm{J}_{a_{5}b_{5}}\bm{P}_{c}\right\rangle _{\mathfrak{%
osp}}, \\
\left\langle \bm{J}_{a_{1}b_{1}}\cdots \bm{J}_{a_{5}b_{5}}\bm{Z}%
_{a_{6}b_{6}}\right\rangle _{\text{M}}& =\alpha _{2}\left\langle \bm{J}%
_{a_{1}b_{1}}\cdots \bm{J}_{a_{6}b_{6}}\right\rangle _{\mathfrak{osp}}, \\
\left\langle \bm{J}_{a_{1}b_{1}}\cdots \bm{J}_{a_{5}b_{5}}\bm{Z}%
_{c_{1}\cdots c_{5}}\right\rangle _{\text{M}}& =\alpha _{2}\left\langle %
\bm{J}_{a_{1}b_{1}}\cdots \bm{J}_{a_{5}b_{5}}\bm{Z}_{c_{1}\cdots
c_{5}}\right\rangle _{\mathfrak{osp}}, \\
\left\langle \bm{QJ}_{a_{1}b_{1}}\cdots \bm{J}_{a_{4}b_{4}}\bar{\bm{Q}}%
\right\rangle _{\text{M}}& =\alpha _{2}\left\langle \bm{QJ}%
_{a_{1}b_{1}}\cdots \bm{J}_{a_{4}b_{4}}\bar{\bm{Q}}\right\rangle _{\mathfrak{%
osp}},  \label{itmalg5}
\end{align}
where $\alpha _{0}$ and $\alpha _{2}$ are arbitrary constants.

It is noteworthy that this invariant tensor for the M~algebra, even if it possesses many more nonzero terms than the supertrace [which would consist of~(\ref{itmalg1}) alone], still misses a lot of other terms present in that for \osp. This is a common feature of $0_{S}$-reduced algebras. In stark contrast, $S$-expanded algebras which do not arise from a $0_{S}$-reduction process do have invariant tensors larger than the one for the original algebra. This fact shapes the dynamics of the theory to a great extent, as we shall see in section~\ref{dyna}.

The supersymmetrized supertrace will be used to provide an invariant tensor
for \osp, with the $32 \times 32$ Dirac
matrices in eleven dimensions as a matrix representation for the bosonic
subalgebra, $\mathfrak{sp}\left(  32\right)  $. The representation with
$\Gamma_{1} \cdots \Gamma_{11} = + \openone$ was chosen. In order to write the Lagrangian, field equations and boundary conditions, it is very useful to have the components of the \osp-invariant tensor with its indices contracted with arbitrary tensors. An explicit calculation gives us
\begin{align}
& L_{1}^{a_{1} b_{1}} \cdots L_{5}^{a_{5} b_{5}} B_{1}^{c} \left\langle \bm{J}_{a_{1} b_{1}} \cdots \bm{J}_{a_{5} b_{5}} \bm{P}_{c} \right\rangle_{\mathfrak{osp}} = \nonumber \\
& \frac{1}{2} \varepsilon_{a_{1} \cdots a_{11}} L_{1}^{a_{1} a_{2}} \cdots L_{5}^{a_{9} a_{10}} B_{1}^{a_{11}},
\label{L5B1}
\end{align}
\begin{align}
& L_{1}^{a_{1}b_{1}}\cdots L_{6}^{a_{6}b_{6}}\left\langle \bm{J}_{a_{1}b_{1}}\cdots\bm{J}_{a_{6}b_{6}}\right\rangle _{\mathfrak{osp}} = 
\frac{1}{3}\sum_{\sigma \in S_{6}} \nonumber \\
& \left[  \frac{1}{4}\Tr\left(  L_{\sigma\left(  1\right)
}L_{\sigma\left(  2\right)  }\right)  \Tr\left(  L_{\sigma
\left(  3\right)  }L_{\sigma\left(  4\right)  }\right)  \Tr%
\left(  L_{\sigma\left(  5\right)  }L_{\sigma\left(  6\right)  }\right)
+\right.  \nonumber\\
& -\Tr\left(  L_{\sigma\left(  1\right)  }L_{\sigma\left(
2\right)  }L_{\sigma\left(  3\right)  }L_{\sigma\left(  4\right)  }\right)
\Tr\left(  L_{\sigma\left(  5\right)  }L_{\sigma\left(
6\right)  }\right)  +\nonumber\\
& \left.  +\frac{16}{15}\Tr\left(  L_{\sigma\left(  1\right)
}L_{\sigma\left(  2\right)  }L_{\sigma\left(  3\right)  }L_{\sigma\left(
4\right)  }L_{\sigma\left(  5\right)  }L_{\sigma\left(  6\right)  }\right)
\right]  ,\label{L6}%
\end{align}
\begin{align}
& L_{1}^{a_{1}b_{1}}\cdots L_{5}^{a_{5}b_{5}}B_{5}^{c_{1}\cdots c_{5}%
}\left\langle \bm{J}_{a_{1}b_{1}}\cdots\bm{J}_{a_{5}b_{5}%
}\bm{Z}_{c_{1}\cdots c_{5}}\right\rangle _{\mathfrak{osp}} = \nonumber\\
& \frac{1}{3}\varepsilon_{a_{1}\cdots a_{11}}\sum_{\sigma \in S_{5}}
\left[  -\frac{5}{4}L_{\sigma\left(  1\right)  }^{a_{1}a_{2}}\cdots
L_{\sigma\left(  4\right)  }^{a_{7}a_{8}}\left[  L_{\sigma\left(  5\right)
}\right]  _{bc}B_{5}^{bca_{9}a_{10}a_{11}}+\right.  \nonumber\\
& +10L_{\sigma\left(  1\right)  }^{a_{1}a_{2}}L_{\sigma\left(  2\right)
}^{a_{3}a_{4}}L_{\sigma\left(  3\right)  }^{a_{5}a_{6}}\left[  L_{\sigma
\left(  4\right)  }\right]  _{\phantom{a}b}^{a_{7}}\left[  L_{\sigma\left(  5\right)
}\right]  _{\phantom{a}c}^{a_{8}}B_{5}^{bca_{9}a_{10}a_{11}}+\nonumber\\
& +\frac{1}{4}L_{\sigma\left(  1\right)  }^{a_{1}a_{2}}L_{\sigma\left(
2\right)  }^{a_{3}a_{4}}L_{\sigma\left(  3\right)  }^{a_{5}a_{6}}B_{5}%
^{a_{7}\cdots a_{11}}\Tr\left(  L_{\sigma\left(  4\right)
}L_{\sigma\left(  5\right)  }\right)  +\nonumber\\
& \left.  -L_{\sigma\left(  1\right)  }^{a_{1}a_{2}}L_{\sigma\left(
2\right)  }^{a_{3}a_{4}}\left[  L_{\sigma\left(  3\right)  }L_{\sigma\left(
4\right)  }L_{\sigma\left(  5\right)  }\right]  ^{a_{5}a_{6}}B_{5}%
^{a_{7}\cdots a_{11}}\right]  ,\label{L5B5}%
\end{align}
\begin{align}
& L_{1}^{a_{1}b_{1}} \cdots L_{4}^{a_{4}b_{4}}\bar{\chi}_{\alpha}\zeta^{\beta
} \left\langle \bm{Q}^{\alpha}\bm{J}_{a_{1}b_{1}}%
\cdots\bm{J}_{a_{4}b_{4}}\bar{\bm{Q}}_{\beta}\right\rangle
_{\mathfrak{osp}} = \nonumber\\
& -\frac{1}{240}\varepsilon_{a_{1}\cdots a_{8}abc}L_{1}^{a_{1}a_{2}}\cdots
L_{4}^{a_{7}a_{8}}\bar{\chi}\Gamma^{abc}\zeta+\nonumber\\
& +\frac{1}{60}\sum_{\sigma \in S_{4}}
\left[  \frac{3}{4}\Tr\left(  L_{\sigma\left(  1\right)
}L_{\sigma\left(  2\right)  }\right)  L_{\sigma\left(  3\right)  }^{a_{1}%
a_{2}}L_{\sigma\left(  4\right)  }^{a_{3}a_{4}}\bar{\chi}\Gamma_{a_{1}\cdots
a_{4}}\zeta+\right.  \nonumber\\
& -2L_{\sigma\left(  1\right)  }^{a_{1}a_{2}}\left[  L_{\sigma\left(
2\right)  }L_{\sigma\left(  3\right)  }L_{\sigma\left(  4\right)  }\right]
^{a_{3}a_{4}}\bar{\chi}\Gamma_{a_{1}\cdots a_{4}}\zeta+\nonumber\\
& +\frac{3}{4}\Tr\left(  L_{\sigma\left(  1\right)  }%
L_{\sigma\left(  2\right)  }\right)  \Tr\left(  L_{\sigma
\left(  3\right)  }L_{\sigma\left(  4\right)  }\right)  \bar{\chi}%
\zeta+\nonumber\\
& \left.  -\Tr\left(  L_{\sigma\left(  1\right)  }%
L_{\sigma\left(  2\right)  }L_{\sigma\left(  3\right)  }L_{\sigma\left(
4\right)  }\right)  \bar{\chi}\zeta\right]  ,\label{L4FF}
\end{align}
where $\Tr$ stands for the trace in the Lorentz indices, i.e.
$\Tr \left( L_{i} L_{j} \right) =
\left( L_{i} \right)_{\phantom{a}b}^{a} \left( L_{j} \right)_{\phantom{b}a}^{b}$.

\section{\label{malglag}The M-Algebra Lagrangian}

We consider a gauge theory on an orientable $\left( 2n+1 \right)$-dimensional manifold $M$ defined by the action
\begin{equation}
S_{\text{T}}^{\left( 2n+1 \right) } \left[ \bm{A}, \bar{\bm{A}} \right] =
\int_{M} L_{\text{T}}^{\left( 2n+1 \right) } \left( \bm{A}, \bar{\bm{A}} \right),
\label{st}
\end{equation}
with the Lagrangian
\begin{align}
L_{\text{T}}^{\left( 2n+1 \right) } \left( \bm{A}, \bar{\bm{A}} \right)
& = k Q_{\bm{A} \leftarrow \bar{\bm{A}}}^{\left( 2n+1 \right)} \nonumber \\
& = \left( n+1 \right) k \int_{0}^{1} dt \left\langle \bm{\theta F}_{t}^{n} \right\rangle.
\label{lag}
\end{align}
Here $\bm{A}$ denotes an M~algebra-valued, one-form gauge connection
\begin{equation}
\bm{A} = \bm{\omega} + \bm{e} + \bm{b}_{2} + \bm{b}_{5} + \bar{\bm{\psi}},
\label{Aone}
\end{equation}
and similarly for $\bar{\bm{A}}$. In Eq.~(\ref{Aone}) each term takes values on a different subspace of the M~algebra, namely
\begin{align}
\bm{\omega}  &  =\frac{1}{2}\omega^{ab}\bm{J}_{ab},\\
\bm{e}  &  =e^{a}\bm{P}_{a},\\
\bm{b}_{2}  &  =\frac{1}{2}b_{2}^{ab}\bm{Z}_{ab},\\
\bm{b}_{5}  &  =\frac{1}{5!}b_{5}^{abcde}\bm{Z}_{abcde},\\
\bar{\bm{\psi}}  &  =\bar{\psi}_{\alpha}\bm{Q}^{\alpha}.
\end{align}

In Eq.~(\ref{st}), $k$ is an arbitrary constant,
$\bm{\theta} = \bm{A} - \bar{\bm{A}}$,
$\bm{A}_{t} = \bar{\bm{A}} + t \bm{\theta}$, and
$\bm{F}_{t} = \mathrm{d} \bm{A}_{t} + \bm{A}_{t}^{2}$.
The Lagrangian~(\ref{lag}) corresponds to a transgression form~\cite{Polaco1,Polaco2,Nosotros,Mora1,CECS,Nosotros2}.
Transgression forms are intimately related to CS forms, since they can be written as the difference of two CS forms plus a boundary term. The presence of this crucial boundary term cures some pathologies present in standard CS Theory, such as ill-defined conserved charges (see Ref.~\cite{CECS}).

The general form of the Lagrangian given in Eq.~(\ref{lag}) suffices in order to derive field equations, boundary conditions and Noether charges. Nevertheless, an explicit version is highly desirable because it clearly shows the physical content of the theory; in particular, a separation in bulk and boundary contributions is essential. This important task can be painstakingly long if approached na\"{\i}vely, i.e. through the sole use of Leibniz's rule. A way out of the bog is provided by the subspace separation method presented in Refs.~\cite{Nosotros,Nosotros2}. This method serves a double purpose; on one hand, it splits the Lagrangian in bulk and boundary terms and, on the other, it allows the separation of the bulk Lagrangian in reflection of the algebra's subspace structure. The method is based on the iterative use of the `Triangle Equation'
\begin{equation}
Q_{\bm{A}\leftarrow\bar{\bm{A}}}^{\left(  2n+1\right)
}=Q_{\bm{A}\leftarrow\tilde{\bm{A}}}^{\left(  2n+1\right)
}+Q_{\tilde{\bm{A}}\leftarrow\bar{\bm{A}}}^{\left(
2n+1\right)  }+\text{d}Q_{\bm{A}\leftarrow\tilde{\bm{A}%
}\leftarrow\bar{\bm{A}}}^{\left(  2n\right)  },
\label{treq}
\end{equation}
Eq.~(\ref{treq}) expresses a transgression form $Q_{\bm{A} \leftarrow \bar{\bm{A}}}^{\left( 2n+1 \right)}$ as the sum of two transgression forms depending on an arbitrary one-form $\tilde{\bm{A}}$ plus a total derivative. This last term has the form
\begin{align}
& Q_{\bm{A} \leftarrow \tilde{\bm{A}} \leftarrow \bar{\bm{A}}}^{\left( 2n \right)} \equiv \nonumber \\
& n \left( n+1 \right) \int_{0}^{1} dt \int_{0}^{t} ds \left\langle \left( \bm{A} - \tilde{\bm{A}} \right) \left( \tilde{\bm{A}} - \bar{\bm{A}} \right) \bm{F}_{st}^{n-1} \right\rangle,
\label{q3}
\end{align}
where
\begin{align}
\bm{A}_{st} &  =\bar{\bm{A}}+s\left(  \bm{A}%
-\tilde{\bm{A}}\right)  +t\left(  \tilde{\bm{A}}%
-\bar{\bm{A}}\right)  ,\label{Ast}\\
\bm{F}_{st} &  =\text{d}\bm{A}_{st}+\bm{A}_{st}%
^{2}.\label{Fst}%
\end{align}

A first splitting of the Lagrangian~(\ref{lag}) is achieved by introducing the intermediate connection $\tilde{\bm{A}}=\bar{\bm{\omega}}$,
\begin{equation}
L\left(  \bm{A},\bar{\bm{A}}\right)  =Q_{\bm{A}%
\leftarrow\bar{\bm{\omega}}}^{\left(  11\right)  }+Q_{\bar
{\bm{\omega}}\leftarrow\bar{\bm{A}}}^{\left(  11\right)
}+\text{d}Q_{\bm{A}\leftarrow\bar{\bm{\omega}}\leftarrow
\bar{\bm{A}}}^{\left(  10\right)  },
\end{equation}
and a second one by separating $Q_{\bm{A} \leftarrow \bar{\bm{\omega}}}^{\left( 11 \right)}$ through $\bm{\omega}$:
\begin{equation}
Q_{\bm{A}\leftarrow\bar{\bm{\omega}}}^{\left(  11\right)
}=Q_{\bm{A}\leftarrow\bm{\omega}}^{\left(  11\right)
}+Q_{\bm{\omega}\leftarrow\bar{\bm{\omega}}}^{\left(
11\right)  }+\text{d}Q_{\bm{A}\leftarrow\bm{\omega
}\leftarrow\bar{\bm{\omega}}}^{\left(  10\right)  }.
\end{equation}

After these two splittings, the Lagrangian~(\ref{lag}) reads
\begin{equation}
L\left(  \bm{A},\bar{\bm{A}}\right)  =Q_{\bm{A}%
\leftarrow\bm{\omega}}^{\left(  11\right)  }-Q_{\bar{\bm{A}%
}\leftarrow\bar{\bm{\omega}}}^{\left(  11\right)  }%
+Q_{\bm{\omega}\leftarrow\bar{\bm{\omega}}}^{\left(
11\right)  }+\text{d}B^{\left(  10\right)  }, \label{lagq3d2}%
\end{equation}
with
\begin{equation}
B^{\left(  10\right)  }=Q_{\bm{A}\leftarrow\bm{\omega
}\leftarrow\bar{\bm{\omega}}}^{\left(  10\right)  }+Q_{\bm{A}%
\leftarrow\bar{\bm{\omega}}\leftarrow\bar{\bm{A}}}^{\left(
10\right)  }. \label{borde}%
\end{equation}
The first two terms in~(\ref{lagq3d2}) are identical (with the obvious
replacements), and we shall mainly concentrate on analyzing them. The third
term will be shown to be unrelated to the two former; in particular, it can be made to vanish without affecting the rest. The boundary term~(\ref{borde}) can be written in a more explicit way by going back to eq.~(\ref{q3}) and replacing the relevant
connections and curvatures. The result is however not particularly
illuminating and, as its explicit form is not needed in order to write
boundary conditions, we shall not elaborate any longer on it.

Let us examine the transgression form
$Q_{\bm{A} \leftarrow \bm{\omega}}^{\left(  11\right)  }$. The subspace separation method
can be used again in order to write down a closed expression for it. To this end we introduce the following set of intermediate connections:
\begin{align}
\bm{A}_{0}  &  =\bm{\omega},\\
\bm{A}_{1}  &  =\bm{\omega}+\bm{e},\\
\bm{A}_{2}  &  =\bm{\omega}+\bm{e}+\bm{b}_{2},\\
\bm{A}_{3}  &  =\bm{\omega}+\bm{e}+\bm{b}_{2}+\bm{b}_{5},\\
\bm{A}_{4}  &  =\bm{\omega}+\bm{e}+\bm{b}_{2}+\bm{b}_{5}+\bar{\bm{\psi}}.
\end{align}
The Triangle Equation~(\ref{treq}) allows us to split the transgression
$Q_{\bm{A}_{4}\leftarrow\bm{A}_{0}}^{\left(  11\right)  }$
following the pattern
\begin{align}
Q_{\bm{A}_{4}\leftarrow\bm{A}_{0}}^{\left(  11\right)  }  &
=Q_{\bm{A}_{4}\leftarrow\bm{A}_{3}}^{\left(  11\right)
}+Q_{\bm{A}_{3}\leftarrow\bm{A}_{0}}^{\left(  11\right)
}+\text{d}Q_{\bm{A}_{4}\leftarrow\bm{A}_{3}\leftarrow
\bm{A}_{0}}^{\left(  10\right)  },\\
Q_{\bm{A}_{3}\leftarrow\bm{A}_{0}}^{\left(  11\right)  }  &
=Q_{\bm{A}_{3}\leftarrow\bm{A}_{2}}^{\left(  11\right)
}+Q_{\bm{A}_{2}\leftarrow\bm{A}_{0}}^{\left(  11\right)
}+\text{d}Q_{\bm{A}_{3}\leftarrow\bm{A}_{2}\leftarrow
\bm{A}_{0}}^{\left(  10\right)  },\\
Q_{\bm{A}_{2}\leftarrow\bm{A}_{0}}^{\left(  11\right)  }  &
=Q_{\bm{A}_{2}\leftarrow\bm{A}_{1}}^{\left(  11\right)
}+Q_{\bm{A}_{1}\leftarrow\bm{A}_{0}}^{\left(  11\right)
}+\text{d}Q_{\bm{A}_{2}\leftarrow\bm{A}_{1}\leftarrow
\bm{A}_{0}}^{\left(  10\right)  }.
\end{align}

Proceeding along these lines one arrives at the Lagrangian
\begin{align}
Q_{\bm{A}_{4} \leftarrow \bm{A}_{0}}^{\left( 11 \right)} & =
6 \left[ H_{a} e^{a} + \frac{1}{2} H_{ab} b_{2}^{ab} + \right. \nonumber \\
& \left. + \frac{1}{5!} H_{abcde} b_{5}^{abcde} - \frac{5}{2} \bar{\psi} \mathcal{R} \text{D}_{\omega} \psi \right].
\label{q40}
\end{align}

All three boundary terms that should in principle appear in~(\ref{q40}) cancel due to the very particular properties of the invariant tensor chosen
[cf.~eqs.~(\ref{itmalg1})--(\ref{itmalg5})].

The tensors $H_{a}$, $H_{ab}$, $H_{abcde}$ and $\mathcal{R}$ are defined as
\begin{align}
H_{a} &  \equiv\left\langle \bm{R}^{5}\bm{P}_{a}\right\rangle
_{\text{M}},\label{H1}\\
H_{ab} &  \equiv\left\langle \bm{R}^{5}\bm{Z}_{ab}%
\right\rangle _{\text{M}},\label{H2}\\
H_{abcde} &  \equiv\left\langle \bm{R}^{5}\bm{Z}%
_{abcde}\right\rangle _{\text{M}},\label{H5}\\
\mathcal{R}_{\phantom{\alpha}\beta}^{\alpha} &  \equiv\left\langle \bm{Q}^{\alpha
}\bm{R}^{4}\bar{\bm{Q}}_{\beta}\right\rangle _{\text{M}}. \label{Rcal}
\end{align}
Explicitly using the invariant tensor~(\ref{L5B1})--(\ref{L4FF}) one finds
\begin{align}
H_{a} & = \frac{\alpha_{2}}{64} R_{a}^{\left( 5 \right)}, \label{H1ex} \\
H_{ab} & = \alpha_{2} \left[ \frac{5}{2} \left( R^{4} - \frac{3}{4} R^{2} R^{2} \right) R_{ab} + 5 R^{2} R_{ab}^{3} - 8 R_{ab}^{5} \right], \label{H2ex} \\
H_{abcde} & = - \frac{5}{16} \alpha_{2} \left[ 5 R_{[ab} R_{cde]}^{\left( 4 \right)} + 40 R_{\phantom{f}[a}^{f} R_{\phantom{g}b}^{g} R_{cde]fg}^{\left( 3 \right)} + \right. \nonumber \\
& \left. - R^{2} R_{abcde}^{\left( 3 \right)} + 4 R_{abcdefg}^{\left( 2 \right)} \left( R^{3} \right)^{fg} \right], \label{H5ex} \\
\mathcal{R} & = - \frac{\alpha_{2}}{40} \left\{ \left( R^{4} - \frac{3}{4} R^{2} R^{2} \right) \openone + \frac{1}{96} R_{abc}^{\left( 4 \right)} \Gamma^{abc} + \right. \nonumber \\
& \left. - \frac{3}{4} \left[ R^{2} R^{ab} - \frac{8}{3} \left( R^{3} \right)^{ab} \right] R^{cd} \Gamma_{abcd} \right\}. \label{Rcalex}
\end{align}

Here we have used the shortcuts%
~\footnote{The trace of the product of an odd number of Lorentz curvatures vanishes: $R^{2n+1}=0$.}
\begin{align}
R^{n}  &  =R_{\phantom{a_{1}}a_{2}}^{a_{1}}\cdots R_{\phantom{a_{n}}a_{1}%
}^{a_{n}},\\
R_{ab}^{n}  &  =R_{ac_{1}}R_{\phantom{c_{1}}c_{2}}^{c_{1}}\cdots
R_{\phantom{c_{n-1}}b}^{c_{n-1}},\\
R_{a_{1}\cdots a_{d-2n}}^{\left(  n\right)  }  &  =\varepsilon_{a_{1}\cdots
a_{d-2n}b_{1}\cdots b_{2n}}R^{b_{1}b_{2}}\cdots R^{b_{2n-1}b_{2n}}.
\end{align}

On section~\ref{dyna} we shall comment on the dynamics produced by this Lagrangian; here we may already note that no derivatives of $e^{a}$, $b_{2}^{ab}$ or $b_{5}^{abcde}$ appear. This can be traced back to the particular form of the invariant tensor~(\ref{itmalg1})--(\ref{itmalg5}), which contains no nonzero components of the form $\left\langle \bm{J}^{3} \bm{PZ}_{2} \right\rangle$, etc.

The last contribution to the Lagrangian~(\ref{lagq3d2}) comes from the
$Q_{\bm{\omega} \leftarrow \bar{\bm{\omega}}}$ term. Taking into account the definition of a transgression form and the form of the invariant tensor, it is straightforward to write down the expression
\begin{equation}
Q_{\bm{\omega} \leftarrow \bar{\bm{\omega}}}^{\left( 11 \right)}
= 3 \int_{0}^{1} dt \theta^{ab} L_{ab} \left( t \right) ,
\end{equation}
where
\begin{equation}
L_{ab}\left( t\right) =\left\langle \bm{R}_{t}^{5}\bm{J}%
_{ab}\right\rangle _{\text{M}}
\end{equation}%
and
\begin{align}
\bm{R}_{t} & = \frac{1}{2}\left[ R_{t}\right] ^{ab}\bm{J}%
_{ab}, \\
\left[ R_{t}\right] ^{ab} & = \bar{R}^{ab}+t\text{D}_{\bar{\bm{\omega}}} \theta^{ab}+t^{2}\theta _{\phantom{a}c}^{a}\theta ^{cb}.
\end{align}

An explicit version for $L_{ab}\left( t\right) $ reads
\begin{align}
L_{ab} \left( t \right) & = \alpha _{0} \left[ \frac{5}{2} \left( R_{t}^{4} - \frac{3}{4} R_{t}^{2} R_{t}^{2} \right) \left[ R_{t} \right]_{ab} + \right. \nonumber \\
& \left. + 5 R_{t}^{2} \left[ R_{t} \right]_{ab}^{3} - 8 \left[ R_{t} \right]_{ab}^{5} \right].
\label{Lab(t)}
\end{align}

A few comments are in order. As seen in~(\ref{Lab(t)}),
$Q_{\bm{\omega } \leftarrow \bar{\bm{\omega}}}^{\left( 11\right) }$ is
proportional to $\alpha _{0}$, as opposed to all other terms, which are
proportional to $\alpha _{2}$. This is a direct consequence of the choice of
invariant tensor. Being the only piece in the Lagrangian unrelated to $\alpha _{2}$,
it can be removed by simply picking $\alpha _{0}=0$. This independence also means that
$Q_{\bm{\omega }\leftarrow \bar{\bm{\omega}}}^{\left( 11\right) }$ is by itself invariant under the
M~algebra. This is related to the fact that this term corresponds to the
only surviving component when the supertrace is used to construct the
invariant tensor.

Because of its form, $Q_{\bm{\omega }\leftarrow \bar{\bm{\omega}}}^{\left( 11\right) }$ apparently contains a bulk interaction of the
$\bm{\omega }$ and $\bar{\bm{\omega}}$ fields. This is no
more than an illusion; in order to realize this, it suffices to use the
`Triangle Equation' with the middle connection set to zero,
\begin{equation}
Q_{\bm{\omega }\leftarrow \bar{\bm{\omega}}}^{\left(
11\right) }=Q_{\bm{\omega }\leftarrow 0}^{\left( 11\right) }-Q_{%
\bar{\bm{\omega}}\leftarrow 0}^{\left( 11\right) }+\text{d}Q_{%
\bm{\omega }\leftarrow 0\leftarrow \bar{\bm{\omega}}%
}^{\left( 10\right) }.
\end{equation}%
Here $Q_{\bm{\omega }\leftarrow 0}^{\left( 11\right) }$ and $Q_{%
\bar{\bm{\omega}}\leftarrow 0}^{\left( 11\right) }$ correspond to
two independient CS exotic-gravity Lagrangians and $Q_{\bm{\omega }%
\leftarrow 0\leftarrow \bar{\bm{\omega}}}^{\left( 10\right) }$
corresponds to the boundary piece relating them.

\subsection{\label{relax}Relaxing Coupling Constants}

All results so far have been obtained from the invariant tensor given in eqs.~(\ref{L5B1})--(\ref{L4FF}). This in turn was derived from the supersymmetrized supertrace of the product of six supermatrices representing as many \osp\ generators. In particular, we have used $32\times32$ Dirac Matrices in $d=11$ to represent the bosonic sector, so that the bosonic components of the invariant tensor correspond to their symmetrized trace~\cite{VanProeyen,Ulf}.

Different invariant tensors may be obtained by considering symmetrized products of traces, as in
$\left\langle \bm{F}^{p} \right\rangle \left\langle \bm{F}^{n-p} \right\rangle$. To exhaust all possibilities one must consider the partitions of six (which is the order of the desired invariant tensor). A moment's thought shows that, apart from the already considered $6=6$ partition, only the $6=4+2$ and $6=2+2+2$ cases contribute, as all others identically vanish. We are thus led to consider the following linear combination:
\begin{equation}
\left\langle \cdots \right\rangle_{\text{M}} =
\left\langle \cdots \right\rangle_{6} +
\beta_{4+2} \left\langle \cdots \right\rangle_{4+2} +
\beta_{2+2+2} \left\langle \cdots \right\rangle_{2+2+2}.
\label{invten3}
\end{equation}
(The coefficient in front of $\left\langle \cdots\right\rangle_{6}$ can be normalized to unity without any loss of generality).

The amazing result of performing this exercise is that no new terms appear in the invariant tensor~(\ref{invten3}); rather, the original rigid structure found in~(\ref{L5B1})--(\ref{L4FF}) is relaxed into one which takes into account the new coupling constants $\beta_{4+2}$ and $\beta_{2+2+2}$. Turning these constants on and off one finds that there are several distinct sectors which are by themselves invariant, so that it is perfectly sensible to associate them with different couplings.

The net effect on the Lagrangian~(\ref{q40}) concerns only the explicit
expressions for the tensors defined in~(\ref{H1})--(\ref{Rcal}); the new
versions read
\begin{align}
H_{a} & = \frac{\alpha_{2}}{64} R_{a}^{\left( 5 \right)}, \label{H1re} \\
H_{ab} & = \alpha_{2}\left[ \frac{5}{2} \left( \kappa_{15} R^{4}- \frac{3}{4} \gamma_{5} R^{2} R^{2} \right) R_{ab} + \right. \nonumber \\
& \left. + 5 \kappa_{15} R^{2} R_{ab}^{3} - 8 R_{ab}^{5} \right], \label{H2re}
\\
H_{abcde} & = -\frac{5}{16} \alpha_{2} \left[ 5 R_{[ab} R_{cde]}^{\left( 4 \right)} + 40 R_{\phantom{f}[a}^{f} R_{\phantom{g}b}^{g} R_{cde]fg}^{\left( 3 \right)} + \right. \nonumber \\
& \left. - \kappa_{15} R^{2} R_{abcde}^{\left( 3 \right)} + 4 R_{abcdefg}^{\left( 2 \right)} \left( R^{3} \right)^{fg} \right], \label{H5re}
\\
\mathcal{R} & =
-\frac{\alpha_{2}}{40} \left\{ \left[ \kappa_{3} R^{4} - \frac{3}{4} \left( 5 \gamma_{9} - 4 \right) R^{2} R^{2} \right] \openone +
\right. \nonumber \\
& + \frac{1}{96} R_{abc}^{\left( 4 \right)} \Gamma^{abc} + \nonumber \\
& \left. - \frac{3}{4}\left[ \kappa_{9} R^{2} R^{ab} - \frac{8}{3}\left( R^{3} \right)^{ab} \right] R^{cd} \Gamma_{abcd} \right\}. \label{Rcalre}
\end{align}

The constants $\kappa _{n}$ and $\gamma _{n}$ are not, as it may seem, an
infinite tower of arbitrary coupling constants, but are rather tightly
constrained by the relations
\begin{align}
\kappa _{m} & = 1+\frac{n}{m}\left( \kappa _{n}-1\right) ,  \label{km} \\
\gamma _{m} & = \gamma _{n}+\left( \frac{n}{m}-1\right) \left( \kappa
_{n}-1\right) .  \label{gm}
\end{align}
These two sets of constants replace the above $\beta _{4+2}$ and $\beta
_{2+2+2}$; once a representative from every one of them has been chosen, the
rest is univocally determined by (\ref{km})--(\ref{gm}). In other words,
fixing one particular $\kappa _{n}$ sets the values of all others. Once all $\kappa _{n}$ are fixed, choosing one $\gamma _{n}$ ties together all the
$\gamma$'s.

The original coupling constants $\beta _{4+2}$ and $\beta _{2+2+2}$ can be
expressed in terms of the new $\kappa _{n}$ and $\gamma _{n}$ as~\footnote{Here $\openone$ denotes the $32 \times 32$ identity matrix, whence $\text{Tr} \left( \openone \right) = 32$.}
\begin{align}
\beta _{4+2}& =\frac{1}{\text{Tr}\left( \openone\right) }n\left( \kappa
_{n}-1\right) , \\
\beta _{2+2+2}& =\frac{15}{\left[ \text{Tr}\left( \openone\right) \right]
^{2}}\left( \gamma _{n}-\kappa _{n}\right) .
\end{align}

It is also worth to notice that
\begin{align}
\beta _{4+2}& =0\qquad \Leftrightarrow \qquad \kappa _{n}=1, \\
\beta _{2+2+2}& =0\qquad \Leftrightarrow \qquad \gamma _{n}=\kappa _{n}.
\end{align}

\subsection{Comparison between the $S$-Expansion Lagrangian and the HTZ Lagrangian}


In Ref.~\cite{Has05}, an action for an eleven-dimensional gauge theory for the M~algebra was found through the Noether procedure. The corresponding Lagrangian can be cast in the form
\begin{equation}
L_{\alpha} = G_{a}e^{a} + \frac{1}{2} G_{ab} b_{2}^{ab} + \frac{1}{5!} G_{abcde} b_{5}^{abcde} - \frac{5}{2} \bar{\psi} \mathcal{Q} \mathrm{D}_{\omega} \psi,
\label{comp2}
\end{equation}
where
\begin{align}
G_{a} & = R_{a}^{\left(  5\right)  }, \label{comp3} \\
G_{ab} & = - 32 \left( 1-\alpha \right) \left[ \left( R^{4} - 2 R^{2} R^{2} \right) R_{ab} + \right. \nonumber \\
& \left. + 5 R^{2} R_{ab}^{3} - 4 R_{ab}^{5} \right], \label{comp4} \\
G_{abcde} & = - \frac{5}{16} \left( 64 \alpha \right) R_{[ab} R_{cde]}^{\left( 4 \right)}, \label{comp5} \\
\mathcal{Q} & = \frac{64}{5} \left[ \frac{1}{96} R^{\left( 4 \right)}_{abc} \Gamma^{abc} + \right. \nonumber \\
& \left. - \frac{1}{2} \left( 1-\alpha \right) \left( R^{2} R_{ab} - R_{ab}^{3} \right) R_{cd} \Gamma^{abcd} \right]. \label{comp6}
\end{align}
Here $\alpha$ is an arbitrary constant.

In our work we have obtained the Lagrangian~(\ref{q40}),
\begin{equation}
L = H_{a} e^{a} + \frac{1}{2} H_{ab} b_{2}^{ab} + \frac{1}{5!} H_{abcde} b_{5}^{abcde} - \frac{5}{2} \bar{\psi} \mathcal{R} \mathrm{D}_{\omega} \psi,
\label{comp7}
\end{equation}
where $H_{a}$, $H_{ab}$, $H_{abcde}$ and $\mathcal{R}$ are given in Eqs.~(\ref{H1re})--(\ref{Rcalre}).

The advantage of writing both Lagrangians in this way is that it makes easier to compare eq.~(\ref{comp2}) with eq.~(\ref{comp7}) just by matching the coefficients $H_{a}$, $H_{ab}$, $H_{abcde}$ and $\mathcal{R}$ with $G_{a}$, $G_{ab}$, $G_{abcde}$ and $\mathcal{Q}$.

Besides an overall multiplicative constant%
~\footnote{In the Lagrangian~(\ref{comp2}), this overall constant corresponds to $\alpha_{2}$. It proves convenient to set this constant as $\alpha_{2}=64$ in order to ease the comparison, see eqs.~(\ref{H1re}) and (\ref{comp3}).},
the Lagrangian~(\ref{comp2}) possesses two tunable independent constants, $\kappa_{n}$ and $\gamma_{n}$, and the Lagrangian~(\ref{comp7}) posseses just one, $\alpha$. An interesting question is if there is some particular choice of $\kappa$'s and $\gamma$'s which allows us to reobtain the HTZ Lagrangian.
Interestingly, the answer is no. As a matter of fact, it can be seen by simple inspection of the expressions for $H_{abcde}$ and $G_{abcde}$ that in the $S$-expansion Lagrangian new terms arise which cannot be wiped out by a simple choice of the $\kappa$ and $\gamma$ constants. Nevertheless, there are some choices which bring both Lagrangians closer. For example, the identification
\begin{align}
 \kappa_{15}&=\frac{\alpha-1}{5}\\
 \gamma_{5}&=\frac{8}{15}\left(\alpha-1\right)
\end{align}
allows us to identify some terms of $H_{ab}$ with the ones in $G_{ab}$. In the same way, the attempt to match~(\ref{Rcalre}) and~(\ref{comp6}) leads to a system of equations which has a solution under some conditions.

Thus the comparison between the Lagrangians~(\ref{q40}) and~(\ref{comp2}) shows the independence between them. The Lagrangian which arises from the $S$-expansion procedure contains all the terms of the HTZ Lagrangian, along with new terms which cannot be made to vanish by a simple choice of constants.

\section{\label{dyna}Dynamics}

\subsection{\label{feq}Field Equations and Four-Dimensional Dynamics}

The field equations for $\bm{A}$ and $\bar{\bm{A}}$ are
completely analogous, and therefore in this section they will be presented
only for $\bm{A}$. The general expression for the field equations reads
\begin{equation}
\left\langle \bm{F}^{5}\bm{T}_{A}\right\rangle _{\text{M}}=0,
\end{equation}
where $\left\{  \bm{T}_{A},A=1,\ldots,\dim\left(  \mathfrak{g}\right)
\right\}  $ is a basis for the algebra and $\bm{F}$ is the curvature.

The field equations obtained by varying $e^{a}$, $b_{2}^{ab}$, $b_{5}^{a_{1}\cdots a_{5}}$ and $\psi$ are given by
\begin{align}
H_{a} &  =0,\label{HaDeltaVielbein=0}\\
H_{ab} &  =0,\label{DeltaB2}\\
H_{abcde} &  =0,\label{DeltaB5}\\
\mathcal{R}\text{D}_{\omega}\psi &  =0,\label{DeltaPsi}%
\end{align}
where explicit expressions for $H_{a}$, $H_{ab}$, $H_{abcde}$ and $\mathcal{R}$ can be found in Eqs.~(\ref{H1re})--(\ref{Rcalre}). The field equation obtained from varying $\omega^{ab}$ reads
\begin{align}
L_{ab} - 10 \left( \text{D}_{\omega} \bar{\psi} \right) \mathcal{Z}_{ab} \left( \text{D}_{\omega} \psi \right) + & \nonumber \\
+ 5H_{abc} \left( T^{c} + \frac{1}{16} \bar{\psi} \Gamma^{c} \psi \right) + & \nonumber \\
+ \frac{5}{2} H_{abcd} \left( \text{D}_{\omega} b^{cd} - \frac{1}{16} \bar{\psi} \Gamma^{cd} \psi \right) + & \nonumber \\
+ \frac{1}{24} H_{abc_{1} \cdots c_{5}} \left( \text{D}_{\omega} b^{c_{1} \cdots c_{5}} + \frac{1}{16} \bar{\psi} \Gamma^{c_{1} \cdots c_{5}} \psi \right) & = 0 ,\label{DeltaOmega}
\end{align}
where we have defined
\begin{align}
L_{ab} &  \equiv\left\langle \bm{R}^{5}\bm{J}_{ab}%
\right\rangle _{\text{M}},\\
\left(  \mathcal{Z}_{ab}\right)  _{\phantom{\alpha}\beta}^{\alpha} &  \equiv\left\langle
\bm{Q}^{\alpha}\bm{R}^{3}\bm{J}_{ab}\bar
{\bm{Q}}_{\beta}\right\rangle _{\text{M}},\\
H_{abc} &  \equiv\left\langle \bm{R}^{4}\bm{J}_{ab}%
\bm{P}_{c}\right\rangle _{\text{M}},\\
H_{abcd} &  \equiv\left\langle \bm{R}^{4}\bm{J}_{ab}%
\bm{Z}_{cd}\right\rangle _{\text{M}},\\
H_{abcdefg} &  \equiv\left\langle \bm{R}^{4}\bm{J}%
_{ab}\bm{Z}_{cdefg}\right\rangle _{\text{M}}.
\end{align}

Explicit versions for these quantities are found using the invariant tensor~(\ref{L5B1})--(\ref{L4FF}):
\begin{equation}
L_{ab}=\alpha_{0}\left[  \frac{5}{2}\left(  R^{4}-\frac{3}{4}R^{2}%
R^{2}\right)  R_{ab}+5R^{2}R_{ab}^{3}-8R_{ab}^{5}\right]  ,
\end{equation}
\begin{align}
\mathcal{Z}_{ab} & = \frac{\alpha_{2}}{40} \left\{
2 \left( R_{ab}^{3} - \frac{3}{4} R^{2} R_{ab} \right) \openone -
\frac{1}{48} R_{abcde}^{\left( 3 \right)} \Gamma^{cde} + \right. \nonumber \\
& - \frac{3}{4} \left( R_{ab} R^{cd} - \frac{1}{2} R^{2} \delta_{ab}^{cd} \right)
R^{ef} \Gamma_{cdef} + \nonumber \\
& - \left[ \delta_{ab}^{cg} R_{gh} R^{hd} R^{ef} -
R_{\phantom{c}a}^{c} R_{\phantom{d}b}^{d} R^{ef} + \right. \nonumber \\
& \left. \left. + \frac{1}{2} \delta_{ab}^{ef} \left( R^{3} \right)^{cd} \right] \Gamma_{cdef} \right\},
\end{align}
\begin{equation}
H_{abc} = \frac{\alpha_{2}}{32} R_{abc}^{\left( 4 \right)},
\end{equation}
\begin{align}
H_{abcd} & = \alpha_{2} \delta_{ab}^{ef} \delta_{cd}^{gh} \left[
\frac{3}{4} R^{2} R_{ef} R_{gh} - R_{ef}^{3} R_{gh} - R_{ef} R_{gh}^{3} + \right. \nonumber \\
& - \frac{4}{5} \left(
R_{eh} R_{fg}^{3} + R_{eh}^{3} R_{fg} - R_{eh}^{2} R_{fg}^{2} \right) + \nonumber \\
& + \frac{1}{2} R^{2} R_{eh} R_{fg} +
\frac{1}{8} \eta_{\left[ ef \right] \left[ gh \right]}
\left( R^{4} - \frac{3}{4} R^{2} R^{2} \right) + \nonumber \\
& \left. - \eta_{fg} \left( R^{2} R_{eh}^{2} - \frac{8}{5} R_{eh}^{4} \right) \right],
\end{align}
\begin{align}
H_{ab c_{1} \cdots c_{5}} & = \frac{\alpha_{2}}{80} \delta_{c_{1} \cdots c_{5}}^{cdefg} \left[
- \frac{5}{3} R_{abcde}^{\left( 3 \right)} R_{fg}
- \frac{1}{6} R_{ab} R_{cdefg}^{\left( 3 \right)} + \right.
\nonumber \\ &
+ 10 R_{abcdepq}^{\left( 2 \right)} R_{\phantom{p}f}^{p} R_{\phantom{q}g}^{q}
- \frac{2}{3} R_{abcdefgpq}^{\left( 1 \right)} \left( R^{3} \right)^{pq} +
\nonumber \\ &
+ \frac{1}{3} R_{\phantom{p}a}^{p} R_{\phantom{q}b}^{q} R_{cdefgpq}^{\left( 2 \right)}
- \frac{1}{3} R_{\phantom{q}a}^{q} R_{bcdefgp}^{\left( 2 \right)} R_{\phantom{p}q}^{p} +
\nonumber \\ &
+ \frac{1}{4} R^{2} R_{abcdefg}^{\left( 2 \right)}
+ \frac{1}{3} R_{\phantom{q}b}^{q} R_{acdefgp}^{\left( 2 \right)} R_{\phantom{p}q}^{p} +
\nonumber \\ &
- \frac{10}{3} \eta_{ga} R_{bcdep}^{\left( 3 \right)} R_{\phantom{p}f}^{p}
+ \frac{10}{3} \eta_{gb} R_{acdep}^{\left( 3 \right)} R_{\phantom{p}f}^{p} +
\nonumber \\ & \left.
- \frac{5}{24} \eta_{\left[ ab \right] \left[ cd \right]} R_{efg}^{\left( 4 \right)} \right].
\end{align}

They satisfy the relationships
\begin{align}
H_{c} &  =\frac{1}{2}R^{ab}H_{abc},\\
H_{cd} &  =\frac{1}{2}R^{ab}H_{abcd},\\
H_{cdefg} &  =\frac{1}{2}R^{ab}H_{abcdefg},\\
\mathcal{R} & = \frac{1}{2} R^{ab} \mathcal{Z}_{ab} .
\end{align}

The problem of finding a `true vacuum' can be analyzed in a similar way to the Refs.~\cite{Has03,Has05}, leading to some results of the above-mentioned references: it is not possible to reproduce four-dimensional General Relativity because there are too many constraints on the four-dimensional
geometry.~\footnote{For an analysis of a similar situation which arises in five dimensions, see Ref.~\cite{Edel06}.}

There are several ways in which one could deal with this problem; as we will
discuss in the conclusions, the excess of constraints is strongly related to
the semigroup choice made in order to construct the M~algebra and also to the
$0_{S}$-reduction. When other semigroups are chosen, different algebras can
arise which reproduce several features of the M~algebra without having its
`dynamical rigidity'~\cite{NosExp}.

\section{\label{final}Summary and Conclusions}

The construction of a transgression gauge field theory for the M~algebra has been developed through the use of two sets of mathematical tools. The first of these sets was provided in Ref.~\cite{NosExp}, where the procedure of expansion is analyzed using abelian semigroups and non-trace invariant tensors for this kind of algebras are written. The problem of the invariant tensor is far from being a trivial one; as discussed in Ref.~\cite{NosExp}, the $0_{S}$-reduction procedure which was necessary in order to construct the M~algebra from \osp\ also renders the supertrace, usually used as invariant tensor, as almost useless. The other set of tools is related with properties of transgression forms, and especially with the subspace separation method~\cite{Nosotros,Nosotros2}, used in order to write down the Lagrangian
in an explicit way.

From a physical point of view, it is very compelling that, using the methods of `dynamical dimensional reduction' introduced in~\cite{Has03,Has05}, something that looks like a `frozen' version of four-dimensional Einstein--Hilbert gravity with positive cosmological constant is obtained by simply abandoning the prejudice that the vacuum should satisfy $\bm{F}=0$. This dynamics `freezing' is a consequence of the constrained form of the invariant tensor: the M~algebra has \emph{more} generators than \osp, but \emph{less} non-vanishing components on the invariant tensor. For this reason, the equations of motion associated to the variations of $e^{a}$, $b_{2}^{ab}$ and $b_{5}^{a_{1} \cdots a_{5}}$ become simply constraints on the gravitational sector. But the poor form of the invariant tensor is a direct consequence of the $0_{S}$-reduction procedure. As shown in Theorem~7.1 from Ref.~\cite{NosExp}, an invariant tensor for a generic $S$-expanded algebra without $0_{S}$-reduction has more non-vanishing components than its $0_{S}$-reduced counterpart and, in general, even more components than the invariant tensor of the original algebra.

The above considerations make it evident that it would be advisable to avoid the $0_{S}$-reduction. The M~algebra arises as the $0_{S}$-reduction of the resonant subalgebra given by eq.~(\ref{ResonantePreM}). This resonant subalgebra itself looks very much like the M~algebra, in the sense that it has the anticommutator
\begin{equation}
\left\{  \bm{Q},\bar{\bm{Q}}\right\}  =\frac{1}{8}\left(
\Gamma^{a}\bm{P}_{a}-\frac{1}{2}\Gamma^{ab}\bm{Z}_{ab}%
+\frac{1}{5!}\Gamma^{a_{1}\cdots a_{5}}\bm{Z}_{a_{1}\cdots a_{5}%
}\right)  ,
\end{equation}
but it also has an \osp\ subalgebra (spanned by $\lambda_{3}\bm{J}_{ab}$, $\lambda_{3}\bm{P}_{a}$, $\lambda_{3}\bm{Z}_{a_{1}\cdots a_{5}}$ and $\lambda_{3}\bm{Q}$; let us remember that $\lambda_{3}\lambda_{3}=\lambda_{3}$). The `central charges' are no longer abelian; rather, their commutators take values on the $\lambda_{3} \times \osp$ sector.
This algebra has a much bigger tensor than the `normal' M~algebra (see
Theorem~7.1 from Ref.~\cite{NosExp}), and therefore, an `unfrozen' dynamics which has good chances of reproducing four-dimensional Einstein--Hilbert Gravity.

A more elegant algebra choice is also shown in Ref.~\cite{NosExp}.
Replacing the M~algebra's semigroup $S_{\text{E}}^{\left(  2\right)  }$
for the cyclic group $\mathbb{Z}_{4},$ a resonant subalgebra of $\mathbb{Z}_{4} \times \osp$ is obtained. It has very interesting features, like having two fermionic charges, $\bm{Q}$ and $\bm{Q}^{\prime}$ with an M~algebra-like anticommutator
\begin{align}
\left\{ \bm{Q}^{\prime}, \bar{\bm{Q}}^{\prime} \right\} & =
\left\{ \bm{Q}, \bar{\bm{Q}} \right\}
\nonumber \\ & =
\frac{1}{8} \left( \Gamma^{a} \bm{P}_{a}
- \frac{1}{2} \Gamma^{ab} \bm{Z}_{ab}
+ \frac{1}{5!} \Gamma^{a_{1} \cdots a_{5}} \bm{Z}_{a_{1} \cdots a_{5}} \right).
\end{align}
Two sets of AdS boost generators, $\bm{P}_{a}$ and $\bm{P}_{a}^{\prime}$, and two (non-abelian) `M5' generators, $\bm{Z}_{a_{1}\cdots a_{5}}$ and $\bm{Z}_{a_{1}\cdots a_{5}}^{\prime}$, are also present. This doubling in several generators makes it specially suitable to construct a transgression gauge field theory. On the other hand, since $\mathbb{Z}_{4}$ is a discrete group, it does not have a zero element; therefore, it has from the outset very good chances of having unfrozen four-dimensional dynamics. Work regarding this issue will be presented elsewhere.

At this point, it is natural to ask ourselves what the relationship between
this M~algebra or M~algebra-like transgression theories and M~Theory could be. It has been proposed that some CS supergravity theories (see Refs.~\cite{Tro97,Horava,Nastase,Banados}) in eleven dimensions could actually correspond to M~Theory, but the potential relations to standard CJS supergravity and String theory remain unsettled. As already discussed, in order to solve these problems it might be wise to take into account the fact that the M~algebra is but one possible choice within a family of superalgebras. Other members of this family [obtained from \osp\ using different abelian semigroups, for instance] might also play a r\^{o}le in finding a truly fundamental symmetry.

\begin{acknowledgments}
F.~I. and E.~R. wish to thank P.~Minning for having introduced them to so many beautiful topics, especially that of semigroups. They are also grateful to D.~L\"{u}st for his kind hospitality at the Arnold Sommerfeld Center for Theoretical Physics in Munich, where part of this work was done. F.~I. and E.~R. were supported by grants from the German Academic Exchange Service (DAAD) and from the Universidad de Concepci\'{o}n (Chile). P.~S. was supported by FONDECYT Grant 1040624 and by Universidad de Concepci\'{o}n through Semilla Grants 205.011.036-1S and 205.011.037-1S.
\end{acknowledgments}

\end{document}